\documentstyle[aps,eqsecnum,preprint,tighten]{revtex}
\begin{document}

\def\mathbf{\bf}
\draft
\preprint{\vbox{Submitted to {\it Physical Review \bf{C}}\hfill
IFUSP/P-1217\\}}
\title{A RELATIVISTIC THOMAS-FERMI DESCRIPTION OF COLLECTIVE MODES IN 
DROPLETS OF NUCLEAR MATTER}
\author{ C. da Provid\^encia, L. Brito and  J. da Provid\^encia} 
\address{ Centro de F\'{\i}sica Te\'orica, Universidade de Coimbra,
3000 - Coimbra, Portugal}
\author{M. Nielsen}
\address{ Instituto de F\'{\i}sica, Universidade de S\~ao Paulo,\\
Caixa Postal 66318 - 5389-970 S\~ao Paulo, S.P., Brazil}
\author{X. Vi\~nas}
\address{Department ECM, Facultat de F\'{\i}sica, Universitat de Barcelona,\\
Diagonal 647, E-08028 Barcelona, Spain}
\maketitle
\begin{abstract}
Isoscalar collective modes in a relativistic meson-nucleon system are
investigated in the framework of the time-dependent Thomas-Fermi method.
The energies of the collective modes are determined by solving
consistently the dispersion relations and the boundary conditions. The
energy weighted sum rule satisfied by the model allows the identification
of the giant ressonances. 
The percentage of the
energy weighted sum rule exhausted by the collective modes is in agreement
with experimental data, but the energies come too high. 
\end{abstract}
\newpage
\section{Introduction}
\label{sec:intro}
Renormalizable relativistic quantum field theories of  hadronic degrees
of freedom, called quantum hadrodynamics (QHD), have been studied for
some time \cite{wal1,serot}. At the level of the mean-field theory
(MFT) and one-loop approximation, these models have proven to be a
powerful tool for describing the bulk properties of nuclear matter.
The binding energy of nuclear matter in MFT
arises from a strong cancellation between repulsive vector and attractive
scalar potentials. Such potentials are comparable to those suggested
by Dirac phenomenology \cite{ha,wa}, Brueckner calculations \cite{wa},
and finite-density QCD sum rules \cite{md}. Therefore, it is not obvious 
that QHD would be able to reproduce the spectrum of finite nuclei, involving 
energies of the order of tens of MeV. However, it has been shown that it can 
realistically describe densities, single-particle  energies and the 
spectrum of collective excitations of finite systems 
\cite{wal1,serot,furn1,wal2,wal3,fuse}.

Collective modes of a relativistic many-body system are characterized
as poles of the meson propagator. However, in the one-loop
approximation, the meson propagators have also poles at space-like
momenta, which arise from polarization effects of the Dirac sea
\cite{soni,tom,perry,furn2}. While the existence of these poles does
not rule out meson-nucleon field theories as useful descriptions of
nuclear systems at low $q$, it may restrict the range of validity 
of several approximations to these theories.
To avoid this problem, in
this work we will study collective excitations of finite nuclear
systems in a semiclassical approximation to the Walecka model.

In refs. \cite{nos1,nos2} a semiclassical approximation to the Walecka
model was introduced to study collective modes in nuclear matter at 
zero and finite
temperature. It was found that the results obtained are 
compatible with microscopic calculations of the meson propagators
\cite{mat,ho}. We want to generalize this semiclassical approach to
the description of collective modes  of finite nuclei by using
a nuclear fluid-dynamical model \cite{vos1,vos2}, which incorporates
monopole and quadrupole distortions of the Fermi surface. This nuclear
fluid-dynamical model has recently been applied with success to the
description of
temperature effects in collective excitations of finite nuclei 
\cite{joao}.

In ref.\cite{fuse} (which is a generalization of the works presented in
refs.\cite{wal2,wal3}) isovector and isoscalar collective modes were 
calculated in the Walecka model, by introducing local hydrodynamic variables
to describe the nucleon fluids with the assumption of irrotational flow and 
in the limit of large masses for the vector mesons. 
As suggested in ref.\cite{fuse}, we lift these restrictions and 
in this work we calculate   the isoscalar collective modes
in the Walecka model in the framework of the
time-dependent Thomas-Fermi method.

In Sec. II we extend the formalism developed in refs. \cite{vos1,vos2}
to the Walecka model.
Collective modes are described by
allowing the meson-fields and the nucleon densities to acquire a time 
dependence. The nucleon motion modifies the source terms in the meson
field equations producing corresponding time-dependent changes in the
meson fields. Since the nucleon dynamics is in turn specified by
the meson fields, collective modes of nuclear motion arise naturally in 
this approach. In Sec. III we derive the equations of motion, boundary 
conditions and orthogonality relations that the normal modes must
satisfy. The dispersion relations,
which  solved consistently
with the boundary conditions,  determine the eigenvalues, are
presented in Sec. IV. In this section  the sum rule satisfied
 by the model is also given.
We identify two rather collective monopole modes at 28 MeV and 35 MeV. 
These
large values are expected since the isoscalar monopole excitation is
a compression mode and, therefore, its energy is related to the
compressibility of nuclear matter \cite{bla}, which is known to be
too low in the Walecka model. For the other multipolarities, we also 
observe that the most collective states come at higher energies than 
the experimentally observed giant resonances. It is true that our lowest 
modes coincide with the modes obtained by \cite{fuse}. 
However, these modes only carry a small 
percentage of the energy 
weighted sum rule and therefore should not be identified 
with the giant resonances.
Finally, in Sec. V we give our numerical results and conclusions.

\section{Fluid-Dynamical Model}
\label{sec:line}

In a classical approximation to the Walecka model the energy of 
a nuclear system
is given by \cite{nos1}
\begin{eqnarray} 
E &=& 4\int\frac{d^3x d^3p}{(2\pi)^3}\,f({\mathbf x},{\mathbf p},t)\,\{[({\mathbf p}
-g_v{\mathbf V})^2 + (M-g_s\sigma)^2]^{1/2} + g_v V_0\}
\nonumber\\*[7.2pt]
&+& \;\frac{1}{2}\int d^3x\,(\Pi_\sigma^2\,+\,{\mathbf\nabla}\sigma\cdot
{\mathbf\nabla}\sigma\,+\,m_s^2\sigma^2) 
\nonumber\\*[7.2pt]
&+& \frac{1}{2}\int d^3x\,[\Pi_{V_i}^2\,-\,2\Pi_{V_i}
\partial_i V_0\,+\,{\mathbf\nabla} V_i\cdot{\mathbf\nabla} V_i\,-\,
\partial_j V_i\partial_i V_j\,+\,m_v^2({\mathbf V}^2 - V_0^2)] \; ,
\label{energy}   
\end{eqnarray}  
where the distribution function, $f({\mathbf x},{\mathbf p},t)$, is restricted
by the requirements
\begin{equation}
N = 4\int {d^3x d^3p\over (2\pi)^3} f({\mathbf x},{\mathbf p},t)\; ,
\end{equation}
\begin{equation}
f^2({\mathbf x},{\mathbf p},t)-f({\mathbf x},{\mathbf p},t)=0\; ,
\end{equation}
and its time evolution is described by the Vlasov equation
\begin{equation}
{\partial f\over\partial t} + \{f,h\}  = 0  \; ,
\label{vla}   
\end{equation}
where $h=\sqrt{({\mathbf p}-g_v{\mathbf V})^2+(M-g_s\sigma)^2}+g_v V_0 =
\epsilon + g_v V_0$ is the classical one-body Hamiltonian and \{,\} denote 
the Poisson Brackets.

The time evolution of the fields is given by
\begin{mathletters}
\begin{eqnarray}
\frac{\partial^2\sigma}{\partial t^2} - \nabla^2\sigma +m_s^2\sigma& =&
g_s\rho_s({\mathbf x},t) \; , \label{em1} 
\\
\frac{\partial^2 V_0}{\partial t^2} - \nabla^2 V_0 + m_v^2 V_0\, &=&
g_v \rho_B({\mathbf x},t) + \frac{\partial}{\partial t}\left( \frac{\partial V_0}
{\partial t} + {\mathbf\nabla}\cdot{\mathbf V}\right ) \; ,\label{em2}  
\\
\frac{\partial^2 V_i}{\partial t^2} - \nabla^2 V_i + m_v^2 V_i\, &=&
g_v j_i({\mathbf x},t) + \frac{\partial}{\partial x_i}\left( \frac{\partial V_0}
{\partial t} + {\mathbf\nabla}\cdot{\mathbf V}\right) \; ,  
\label{em3}
\end{eqnarray}
\end{mathletters}
with
\begin{eqnarray}
\rho_s({\mathbf x},t)&=&4\int\frac{d^3p}{(2\pi)^3}\,f({\mathbf x},{\mathbf p},t)\,
\frac{M-g_s\sigma}{\epsilon} \; , 
\\
\rho_B({\mathbf x},t)&=&4\int\frac{d^3p}{(2\pi)^3}\,f({\mathbf x},{\mathbf p},t)
 \; , 
\\
{\mathbf j}({\mathbf x},t)&=&4\int\frac{d^3p}{(2\pi)^3}\,f({\mathbf x},{\mathbf p},t)\,
\frac{{\mathbf p}-g_v{\mathbf V}}{\epsilon} \; .
\label{fontes}
\end{eqnarray}

Using the Vlasov equation, Eq.(\ref{vla}), it can be easily shown that
the four-current satisfies the continuity equation, and that the 
components of the vector field are related through \cite{nos1}: 
\begin{equation}
 \partial_\mu V^\mu = 0   \; .
\end{equation}
Therefore, the second term on the right-hand side of 
Eqs.(\ref{em2},\ref{em3}) vanishes.

In our calculations we will assume that the density of a spherical 
nucleus in the ground-state  is constant inside the nucleus and
zero outside, and is given by 
\begin{equation}
\rho_0(r)=
4\int\frac{d^3p}{(2\pi)^3}\,f_0({\mathbf x},{\mathbf p})\; .
\label{eqq} 
\end{equation}
with
\begin{equation}
f_0({\mathbf x},{\mathbf p}) = \Theta[p_F^2(r)-{\mathbf p}^2]\; ,
\label{eq} 
\end{equation}
where $p_F(r)=\overline{p}_F\Theta[R_0-r]$, $\overline{p}_F$ is the 
nuclear matter Fermi momentum, and   $R_0$ is the nuclear 
radius.
The ground-state distribution function
$f_0$ is determined by the particle number $A$ and by the minimization of the 
energy and the  equilibrium nuclear
matter density, $\overline\rho_0$, is calculated from
 equations (\ref{eq}) and (\ref{eqq})
$$\rho_0(r)=\overline\rho_0\Theta[R_0-r].$$

Giant resonances manifest themselves as small amplitude highly collective modes.
Therefore, they are described at the  microscopic level  by the
 RPA equations. In the classical limit, these equations are
obtained by the linearization of the Vlasov equation. In this
context we begin by expanding the distribution function around its
equilibrium value $f_0({\mathbf x},{\mathbf p})$:
\begin{equation}
f({\mathbf x},{\mathbf p},t)= f_0({\mathbf x},{\mathbf p})+\{S,f_0\}
+{1\over2}\{S,\{S,f_0\}\}+...  \; ,
\label{f}
\end{equation}
where $S({\mathbf x},{\mathbf p},t)$ is a generating function which 
describes small deviations from equilibrium.

In its more general form, the distribution function, $f({\mathbf x},
{\mathbf p},t)$, should include static as well as dynamic deformations
of the nuclear system. For this reason we decompose the infinitesimal
generator $S({\mathbf x},{\mathbf p},t)$ into a time-even and a time-odd part
\begin{mathletters}
\begin{eqnarray}
S({\mathbf x},{\mathbf p},t)&=&P({\mathbf x},{\mathbf p},t)+Q({\mathbf x},{\mathbf p},t)\; ,
\\
Q({\mathbf x},{\mathbf p},t)&=&Q({\mathbf x},-{\mathbf p},t)\; ,
\\
P({\mathbf x},{\mathbf p},t)&=&-P({\mathbf x},-{\mathbf p},t)\; .
\end{eqnarray}
\end{mathletters}
The  time-even generator, $Q({\mathbf x},{\mathbf p},t)$, takes into account
the dynamic deformations. The static deformations are
described by the time-even distribution function, which includes
the fields responsible for the deformations of the Fermi surface. In the
present approach, it is
expressed in terms of the time-odd generator $P({\mathbf x},{\mathbf p},t)$ 
\begin{eqnarray}
f_E({\mathbf x},{\mathbf p},t)&=& f_0({\mathbf x},{\mathbf p})+\{P,f_0\}
+{1\over2}\{P,\{P,f_0\}\}+...  
\nonumber\\*[7.2pt]
&=&\Theta[\lambda-h_0({\mathbf x},{\mathbf p})-W({\mathbf x},t)-
{1\over2}p_ip_j\chi_{ij}({\mathbf x},t)] \; .
\label{fe}
\end{eqnarray}
The scalar field, $W({\mathbf x},t)$, is related to the deformations
which preserve the spherical form of the Fermi surface. The tensor field,
$\chi_{ij}({\mathbf x},t)$, introduces deformations in the Fermi
sphere. Hopefully, the scalar and tensor fields will
provide an adequate description of the monopole and quadrupode deformations
of the Fermi sphere. In Eq.(\ref{fe}), $h_0({\mathbf x},{\mathbf p})
=\sqrt{{\mathbf p}^2+{M^*}^2({\mathbf x})}+g_v V_0^0({\mathbf x})$, with
$M^*({\mathbf x})=M-\sigma_0({\mathbf x})$, and $\sigma_0({\mathbf x})$
and $V_0^0({\mathbf x})$
are, respectively,  
the equilibrium values of the
fields $\sigma$ and $V_0$. The Fermi momentum is related to $\lambda$ through
\begin{equation}
\lambda=\sqrt{p_F^2(r)+{M^*}^2({\mathbf x})}+g_v V_0^0({\mathbf x})=\epsilon_F
+g_v V_0^0({\mathbf x})   \; .
\end{equation}

The introduction of the generator $Q({\mathbf x},{\mathbf p},t)$ destroys the 
time reflexion invariance of the equilibrium distribution function. It will 
allow for the appearance of transverse flow \cite{st} in the nucleus.
The simplest choice which includes this possibility is given by
\cite{vos2}
\begin{equation}
Q({\mathbf x},{\mathbf p},t)=\psi({\mathbf x},t)+{1\over 2}p_ip_j\phi_{ij}
({\mathbf x},t) \; ,
\label{fo}
\end{equation}
where $\psi({\mathbf x},t)$ and $\phi_{ij}({\mathbf x},t)$ are, respectively,  
scalar and symmetrical tensor fields.

The time evolution of the generator $S$ and the field fluctuations
are determined by the appropriate Lagrangian. 
For small deviations from equilibrium it is enough to consider the 
quadratic Lagrangian
\begin{equation}
L^{(2)}=2\int\frac{d^3pd^3x}{(2\pi)^3}\,f_0\{S,\dot{S}\}+
\int d^3x\Pi_\sigma\dot{\sigma}+ \int d^3x \Pi_{V_i}\dot{V}_i 
-E^{(2)} \; .
\label{l2}
\end{equation}
Using the ansatz Eqs.(\ref{f}), (\ref{fe}) and (\ref{fo}), decomposing
the boson fields into a static (ground-state) contribution 
and  a small time-dependent increment and imposing the barion number 
conservation, we get
\begin{eqnarray}
& &{2}\int\frac{d^3pd^3x}{(2\pi)^3}\,f_0\{S,\dot{S}\}=\int d^3x 
\epsilon_F\left[{2p_F\over\pi^2}\left(W+{p_F^2\over6}\chi_{ii}\right)
\left(\dot{\psi}+{p_F^2\over6}\dot{\phi}_{ii}\right)\right. 
\nonumber\\*[7.2pt]
&+&\left.{p_F^2\rho_0\over 10}\left(\chi_{ij}-{\delta_{ij}\over3}
\chi_{kk}\right)\left(\dot{\phi}_{ij}-{\delta_{ij}\over3}
\dot{\phi}_{kk}\right)\right]+
\int d{\mathbf\Sigma}.\delta{\mathbf R}\overline{\rho}_0\left(
\dot{\psi}+{\overline p_F^2\over10}\dot{\phi}_{ii}\right) \; ,
\label{l21}
\end{eqnarray}
\begin{eqnarray}
E^{(2)}&=&\int d^3x\left[{p_F\epsilon_F\over\pi^2}W^2 + {\epsilon_F
\rho_0\over2}W\chi_{ii} + {\rho_0\over2\epsilon_F}{\mathbf\nabla}\psi\cdot
{\mathbf\nabla}\psi + {\epsilon_Fp_F^2\rho_0\over20}\left({\chi^2_{ii}
\over2}+\chi^2_{ij}\right) \right.
\nonumber\\*[7.2pt]
&+&{p_F^2\rho_0\over10\epsilon_F}\left({\mathbf\nabla}\psi\cdot
{\mathbf\nabla}\phi_{ii}
+2\partial_i\psi\partial_j\phi_{ij}\right) +
{p_F^4\rho_0\over280\epsilon_F}\left(4\partial_j\phi_{ii}
\partial_k\phi_{jk}+{\mathbf\nabla}\phi_{ii}\cdot{\mathbf\nabla}\phi_{jj}
\right.
\nonumber\\*[7.2pt]
&+&\left.\left. 
2{\mathbf\nabla}\phi_{ij}\cdot{\mathbf\nabla}\phi_{ij}+4\partial_i\phi_{ij}
\partial_k\phi_{kj}+4\partial_k\phi_{ij}\partial_j\phi_{ik}\right)
\right]+\int d^3x\left[(g_sM^*\delta\sigma\right.
\nonumber\\*[7.2pt]
&-&\left.g_v\epsilon_F\delta V_0)
\left({2p_F\over\pi^2}W+{\rho_0\over2}\chi_{ii}\right) +
{g_v\rho_0\over\epsilon_F}\delta V_j\left(\partial_j\psi+{p_F^2\over
10}(\partial_j\phi_{ii}+2\partial_i\phi_{ij})\right)\right]
\nonumber\\*[7.2pt]
&+& \;\frac{1}{2}\int d^3x\,[\Pi_\sigma^2\,+\,{\mathbf\nabla}\delta\sigma\cdot
{\mathbf\nabla}\delta\sigma\,+\,(m_s^2+\Delta m_s^2)(\delta \sigma)^2]\, +\,
\frac{1}{2}\int d^3x\,[\Pi_{V_i}^2 
\nonumber\\*[7.2pt]
&-&2\Pi_{V_i}\partial_i\delta V_0\,+\,{\mathbf\nabla}\delta V_i\cdot{\mathbf\nabla} 
\delta V_i\,-\,\partial_j\delta V_i\partial_i\delta V_j\,+\,
(m_v^2+\Delta m_v^2)(\delta{\mathbf V})^2 
\nonumber\\*[7.2pt]
&-& m_v^2 (\delta V_0)^2]\, + \,\int d{\mathbf\Sigma}.\delta{\mathbf R}\,
(g_s\overline\rho_{s0}\delta\sigma-g_v\overline{\rho}_0\delta V_0)
\; .
\label{e2}
\end{eqnarray}
The surface integrals in the above equations take into account 
possible surface displacements
parametrized by a vector field, 
$\delta{\mathbf R}({\mathbf x})$. 
Our choice of the even distribution function allows 
explicitly for this effect.
In Eq.(\ref{e2}), $\Delta m_s^2=g_s^2{\partial\rho_{s0}\over\partial M^*}$
and $\Delta m_v^2=g_v^2\rho_0/\epsilon_F$.

\section{Equations of Motion, Boundary Conditions and Orthogonality
Relations}

The equations of motion and boundary conditions that 
specify the dynamics of the fields are obtained from Eq.(\ref{l2})
through the Euler-Lagrange equations. We get
\begin{mathletters}
\begin{equation}
\delta\dot\sigma=\Pi_\sigma\; ,
\label{mo1}
\end{equation}
\begin{equation}
\dot\Pi_\sigma-\nabla^2\delta\sigma+(m_s^2+\Delta m_s^2)\delta\sigma
=-g_s M^* \left({2 p_F\over \pi^2} W
+{\rho_0\over2}\chi_{ii}\right)\; ,
\label{mo2}
\end{equation}
\begin{equation}
\delta\dot V_i = \Pi_{V_i}-\partial_i\delta V_0 \; ,
\label{mo3}
\end{equation}
\begin{equation}
\delta\ddot{V_i}-\nabla^2\delta V_i+ (m_v^2+\Delta m_v^2)\delta
 V_i=-{g_v\over \epsilon_{F}} \rho_0\left(\partial_i\psi+
{p_F^2\over 10}\left(\partial_i\phi_{jj}+2\partial_j\phi_{ij}\right)
\right)\; ,
\label{mo4}
\end{equation}
\begin{equation}
\delta\ddot{V_0}-\nabla^2\delta V_0+ m_v^2\delta V_0=-g_v\epsilon_{F}  
\left({2 p_F\over\pi^2} W+{\rho_0\over2}\chi_{ii}\right)\; ,
\label{mo5}
\end{equation}
\begin{equation}
\dot{W}+{p_F^2\over6}\dot{\chi}_{ii}
={p_F^2\over3\epsilon_{F}^2}\nabla^2\psi+{p_F^4\over30
\epsilon_{F}^2}\left(\nabla^2\phi_{ii}+2\partial_i\partial_j\phi_{ij}
\right)+{g_v p_F^2\over3\epsilon_{F}^2}\partial_i\delta V_i\; ,
\label{mo6}
\end{equation}
\begin{equation}
\dot{\psi}+{p_F^2\over6}\dot{\phi}_{ii}={W}+{p_F^2\over6}{\chi}_{ii}
-\left(g_v\delta V_0-g_s {M^*\over\epsilon_F}\delta\sigma\right)\; ,
\label{mo7}
\end{equation}
\begin{equation}
\left(\dot{\psi}+{p_F^2\over10}\dot{\phi}_{kk}\right)
\delta_{ij}+{p_F^2\over5}\dot{\phi}_{ij}
= {W}\delta_{ij}+{p_F^2\over10}\left({\chi}_{kk}\delta_{ij}
+2{\chi}_{ij}\right)-\delta_{ij}
\left(g_v\delta V_0-g_s{M^*\over \epsilon_F} \delta\sigma\right)\; ,
\label{mo8}
\end{equation}
\begin{eqnarray}
&&\left(\dot{W}+{p_F^2\over10}\dot{\chi}_{kk}\right)\delta_{ij}+
{p_F^2\over5}\dot{\chi}_{ij}  =  {p_F^2\over 5 \epsilon_{F}^2}
\left(\nabla^2 \psi \delta_{ij} + 2\partial_i\partial_j \psi\right)
+{g_v p_F^2\over5\epsilon_{F}^2}
(\partial_k\delta V_k\delta_{ij} + \partial_i\delta V_j
\nonumber\\*[7.2pt]
&+& \partial_j\delta V_i) 
+ {p_F^4\over 35 \epsilon_{F}^2}\left[\delta_{ij}
\left({1\over2}\nabla^2 \phi_{kk}  
+ \partial_k\partial_l \phi_{kl}\right)+\nabla^2 \phi_{ij} +
\partial_i\partial_j \phi_{kk}+2 \partial_i\partial_k \phi_{kj}
+2 \partial_j\partial_k \phi_{ki}\right]\; .
\label{mo9}
\end{eqnarray}
\end{mathletters}
It is worth mentioning that Eqs.(\ref{mo1}) to (\ref{mo9}) are
valid only in the interior of the nucleus. Therefore, we replace 
$p_F$, $\epsilon_F$ and $\rho_0$ in these equations by their equilibrium
values. 
At the surface, the variational fields satisfy  the following boundary conditions
\begin{mathletters}
\begin{equation}
\left.{x}_k(\partial_k\delta\sigma+g_s\overline\rho_{s0}\delta 
R_k)\right|_{r=R_0}=0\; ,\label{bo1}
\end{equation}
\begin{equation}
\left.{x}_k(\partial_k\delta V_i-\partial_i\delta V_k)
\right|_{r=R_0}=0\; ,\label{bo2}
\end{equation}
\begin{equation}
\left.{x}_k(\partial_k\delta V_0+\delta\dot{V}_k+g_v\overline\rho_0
\delta R_k)\right|_{r=R_0}=0\; ,\label{bo3}
\end{equation}
\begin{equation}
\left.{x}_k\left(\partial_k\psi+{\overline p_F^2\over10}
(\partial_k\phi_{ii}+2\partial_i\phi_{ik})+g_v\delta V_k
+\overline\epsilon_F\delta\dot{R}_k\right)
\right|_{r=R_0}=0\; ,\label{bo4}
\end{equation}
\begin{eqnarray}
&&{x}_k\left[\delta\dot{R}_k\delta_{ij}+{1\over\overline\epsilon_F}(
\partial_k\psi\delta_{ij}+\partial_i\psi\delta_{jk}+\partial_j\psi
\delta_{ik})+{\overline p_F^2\over7\overline\epsilon_F}
(\partial_l\phi_{kl}\delta_{ij}\right.
\nonumber\\*[7.2pt]
&+&\partial_j\phi_{ll}\delta_{ik}+\partial_l\phi_{lj}\delta_{ik}+
\partial_l\phi_{li}\delta_{jk}+
{1\over2}\partial_k\phi_{ll}\delta_{ij}+\partial_k\phi_{ij}+
\partial_j\phi_{ik}+\partial_i\phi_{jk})
\nonumber\\*[7.2pt]
&+&\left.\left.{g_v\over\overline\epsilon_F}(\delta V_k\delta_{ij}+
\delta V_i\delta_{jk}+\delta V_j\delta_{ik})-(\xi_i\delta_{jk}
+\xi_j\delta_{ik})\right]\right|_{r=R_0}=0 \; ,\label{bo5}
\end{eqnarray}
\begin{equation}
\left.\dot{\psi}+{\overline p_F^2\over10}\dot{\phi}_{ii}+g_v\delta V_0
-g_s{\overline\rho_{s0}\over\overline\rho_0}\delta\sigma
\right|_{r=R_0}=0\; .\label{bo6}
\end{equation}
\end{mathletters}
In order to ensure that the current density is not singular at the
surface, the following boundary condition has also to be imposed
\cite{vos2}
\begin{equation}
\left.{x}_k\phi_{kj}\right|_{r=R_0}=0 \; .\label{bo7}
\end{equation}
In Eq.(\ref{bo5}), ${\mathbf\xi}$ is a vector Lagrange multiplier that takes
into account the restriction (\ref{bo7}).

We look for normal-mode solutions where all the fields oscillate
harmonically in time. This means that the fields are described by a
superposition of the real parts of $\{\Pi_\sigma^{(n)},\,\psi^{(n)},\,
\delta{\mathbf V}^{(n)},\,\phi_{ij}^{(n)},\,i\delta\sigma^{(n)},\,i\delta
V_0^{(n)},\,iW^{(n)},\,i\delta{\mathbf R}^{(n)},\,i{\mathbf\Pi}_V^{(n)},\,
i\chi_{ij}^{(n)}\}\exp^{-i\omega_n t}$, where all the quantities within 
the braces are only functions of ${\mathbf x}$. This normal-mode analysis
leads to the RPA coupled equations for the eigenmodes:
\begin{mathletters}
\begin{equation}
\omega_n\delta\sigma^{(n)}=\Pi_\sigma^{(n)} \; ,
\label{nm1}
\end{equation}
\begin{equation}
-\omega_n^2\delta{\sigma}^{(n)}-\nabla^2\delta\sigma^{(n)}
+(m_s^2+\Delta m_s^2)\delta\sigma^{(n)}=-g_s M^* \left({2 
\overline{p}_F\over  \pi^2} W^{(n)}
+{\overline{\rho}_0\over2}\chi_{ii}^{(n)}\right)\; ,
\label{nm2}
\end{equation}
\begin{equation}
-\omega_n\delta V_i^{(n)}=\Pi_{V_i}^{(n)}-\partial_i\delta V_0^{(n)}
\; ,\label{nm3}
\end{equation}
\begin{eqnarray}
-\omega_n^2\delta{V_i}^{(n)}&-&\nabla^2\delta V_i^{(n)}
+ (m_v^2+\Delta m_s^2)V_i^{(n)}=-{g_v\over \overline{\epsilon}_{F}} 
\overline{\rho}_0\left(\partial_i\psi^{(n)}\right.
\nonumber\\*[7.2pt]
&+&\left.{\overline{p}_F^2\over 10}
\left(\partial_i\phi_{jj}^{(n)}+2\partial_j\phi_{ij}^{(n)}\right)\right)
\; , \label{nm4}
\end{eqnarray}
\begin{equation}
-\omega_n^2\delta{V_0}^{(n)}-\nabla^2\delta V_0^{(n)}+ m_v^2\delta 
V_0^{(n)}=-\overline{\epsilon}_{F} g_v \left({2 
\overline{p}_F\over \pi^2} 
W^{(n)}+{\overline{\rho}_0\over2}\chi_{ii}^{(n)}\right)\; ,
\label{nm5}
\end{equation}
\begin{equation}
\omega_n\left({W}^{(n)}+{\overline{p}_F^2\over6}{\chi}_{ii}^{(n)}\right)
={\overline{p}_F^2\over3\overline{\epsilon}_{F}^2}\nabla^2\psi^{(n)}+
{\overline{p}_F^4\over30\overline{\epsilon}_{F}^2}\left(\nabla^2
\phi_{ii}^{(n)}+2\partial_i\partial_j\phi_{ij}^{(n)}\right)+
{g_v\overline{p}_F^2\over3\overline{\epsilon}_{F}^2}\partial_i
\delta V_i^{(n)}\; ,
\label{nm6}
\end{equation}
\begin{equation}
-\omega_n\left({\psi}^{(n)}+{\overline{p}_F^2\over6}{\phi}_{ii}
^{(n)}\right) = {W}^{(n)}+{\overline{p}_F^2\over6}{\chi}_{ii}^{(n)}
-\left(g_v\delta V_0^{(n)}-g_s{M^*\over \overline{\epsilon}_F} 
\delta\sigma^{(n)}\right)\; ,
\label{nm7}
\end{equation}
\begin{equation}
-\omega_n\phi_{ij}^{(n)} = \chi_{ij}^{(n)}\;\;\;\;\;\;\;\; 
(i\neq j)\; ,\label{nm8}
\end{equation}
\begin{eqnarray}
\omega_n\chi_{ij}^{(n)}&=&{2\over \overline{\epsilon}_F^2}
\partial_i\partial_j \psi^{(n)}+{\overline{p}_F^2\over7
\overline{\epsilon}_F^2}(\nabla^2 \phi_{ij}^{(n)}+
\partial_i\partial_j \phi_{kk}^{(n)}+2 \partial_i\partial_k 
\phi_{kj}^{(n)}+2 \partial_j\partial_k \phi_{ki}^{(n)})
\nonumber\\*[7.2pt]
&+&{g_v \over\overline{\epsilon}_{F}^2}(\partial_i\delta V_j^{(n)}+
\partial_j\delta V_i^{(n)}) \;\;\;\;\;\;\;\; (i\neq j)\; .
\label{nm9}
\end{eqnarray}
\end{mathletters}
It is clear form Eq.(\ref{nm8}) that $\chi_{ij}$ and $\phi_{ij}$ are
canonically conjugate fields.

The solutions of the  the above equations satisfy the following
orthogonality relation
\begin{eqnarray}
\int d^3x \, \epsilon_F\left[{2 p_F\over\pi^2}\left(W^{(m)}\right.\right.
&+&\left.\left.{p_F^2\over6}\chi_{ii}^{(m)}\right)
\left({\psi^{(n)}}+{p_F^2\over6}{\phi}_{ii}^{(n)}\right)+ 
{p_F^2\rho_0\over 10}\left(\chi_{ij}^{(m)}-{\delta_{ij}\over3}
\chi_{kk}^{(m)}\right)\left({\phi}_{ij}^{(n)}-{\delta_{ij}\over3}
{\phi}_{kk}^{(n)}\right)\right]
\nonumber\\*[7.2pt]
-\int d^3x\Pi_\sigma^{(n)}{\delta\sigma}^{(m)}&+& \int d^3x 
\Pi_{V_i}^{(m)}{\delta V}_i^{(n)} 
+\int d{\mathbf\Sigma}.\delta{\mathbf R}^{(m)}\overline{\rho}_0\left(
{\psi}^{(n)}+{\overline p_F^2\over10}{\phi}_{ii}^{(n)}\right)=-\delta_{mn} 
\; , \label{norma}
\end{eqnarray}

\section{Dispersion Relations and Sum Rules}
\subsection{Dispersion Relations}

The eletric 
modes are described by the same  kind of solutions  as
constructed in ref.\cite{vos2}, {\it i. e.}, by two kinds of transverse
fields
\begin{eqnarray}
\left[\phi_{ij}\right]_1& =& \{(\partial_i\partial_j-\delta_{ij}
\nabla^2)l^2
-[\partial_i({\mathbf\nabla}\times{\mathbf l})_j+\partial_j({\mathbf\nabla}
\times{\mathbf l})_i]
\nonumber\\*[7.2pt]
&-&[({\mathbf\nabla}\times{\mathbf l})_i({\mathbf\nabla}\times{\mathbf l})_j
+({\mathbf\nabla}\times{\mathbf l})_j({\mathbf\nabla}\times
{\mathbf l})_i]\}j_l(k_1r)Y_{l0} \; ,
\label{t1}
\end{eqnarray}
\begin{equation}
\left[\phi_{ij}\right]_2 = [\partial_i({\mathbf\nabla}\times{\mathbf l})_j+
\partial_j({\mathbf\nabla}\times{\mathbf l})_i]j_l(k_2r)Y_{l0} \; ,
\label{t2}
\end{equation}
and by one  longitudinal tensor field
\begin{equation}
\left[\phi_{ij}\right]_3 = \left(\partial_i\partial_j-{\delta_{ij}
\over3}\nabla^2\right)j_l(k_3r)Y_{l0} \; .
\label{t3}
\end{equation}

The advantage of using the above  combination of  the
four linearly independent angular tensor functions: $\partial_i
\partial_jY_{l0}$, $\delta_{ij}Y_{l0}$, $(x_i\partial_j+
x_j\partial_i)Y_{l0}$, and $x_ix_jY_{l0}$, is that all solutions
given above are traceless. In particular, the transverse fields
also verify the relations
\begin{equation}
\partial_i\left[\phi_{ij}\right]_1=0\;\;\;\;\;\;\;\nobreak\,\mbox{and}
\;\;\;\;\;\;\;\partial_i\partial_j\left[\phi_{ij}\right]_2=0\; .
\label{re}
\end{equation}
For each multipolarity, all  scalar fields are proportional
to $j_l(kr)Y_{l0}$, and the vector fields are combinations
of two linearly independent vector functions: $\partial_i
(j_l(kr)Y_{l0})$ and $({\mathbf\nabla}\times{\mathbf l})_ij_l(kr)Y_{l0}$.
Using these combinations in Eqs.(\ref{nm1}) to (\ref{nm9})  
it is straightforward to show that
the transverse solutions do not couple to the scalar fields, and one
has $[\delta\sigma]_{1,2}=[\delta V_0]_{1,2}=[W]_{1,2}=[\psi]_{1,2}=
[\Pi_\sigma]_{1,2}=0$. For solutions of kind 1, the vector fields are
also zero: $[\delta V_i]_1=[\Pi_{V_i}]_1=[\delta R_i]_1=0$ and the
dispersion relation for this particular solution is given by
\begin{equation}
\omega_n^2={\overline{p}_F^2\over7\overline{\epsilon}_F^2}k_1^2 \; .
\label{d1}
\end{equation}
This is the same relation as obtained in ref.\cite{vos2}. This should be
expected since the meson fields, which are the new ingredients in
the model used here, do not couple to the solution of kind 1.

For solutions of kind 2, we still have $[\delta R_i]_2=0$, since, from 
Eq.(\ref{bo1}), the vector field $\delta{\mathbf R}$ is directly related to 
the scalar field $\delta\sigma$. However, the vector fields $[\delta V_i]_2$
and $[\Pi_{V_i}]_2$ are coupled to the tensor fields. We get
\begin{equation}
[\delta V_i]_2={G_v(k_2)\over5g_v}\overline{p}_F^2\partial_j
\left[\phi_{ij}\right]_2=-{G_v(k_2)\over5g_v}\overline{p}_F^2k_2^2
({\mathbf\nabla}\times{\mathbf l})_ij_l(k_2r)Y_{l0}\; ,
\end{equation}
where 
\begin{equation}
G_v(k)={g_v\overline{\rho}_0\over\overline{\epsilon}_F(\omega_n^2
-k^2-{m^*_v}^2)} \; , 
\label{gv}
\end{equation}
and ${m^*_v}^2=m_v^2+\Delta m_v^2$.
Using the solutions of kind 2 in the normal mode equations we get
\begin{equation}
\left(\omega_n^2-{3k_2^2\overline{p}_F^2\over7\overline{\epsilon}_F^2}\right)
(\omega_n^2
-k_2^2-{m^*_v}^2)={g_v^2k_2^2\overline{p}_F^2\overline{\rho}_0
\over5\overline{\epsilon}_F^3} \; ,
\label{d2}
\end{equation}
which give us two different solutions for $k_2^2$. For $g_v=0$, one
of the solutions is exactly the same which is obtained in ref.\cite{vos2}. This
solution is now modified and a new solution appears, due to the coupling 
between the vector meson field and the fields introduced to describe
the nuclear deformations.

The longitudinal  solutions, $\left[\phi_{ij}\right]_3$, couple
to all other fields and give
\begin{mathletters}
\begin{equation}
[\psi]_3=f(k_3)j_l(k_3r)Y_{l0}\; ,
\end{equation}
\begin{equation}
[W]_3=-{\omega_n[\psi]_3\over G_{s0}(k_3)}\; ,
\end{equation}
\begin{equation}
[\delta\sigma]_3={2g_sM^*\overline{p}_F\over\pi^2(\omega_n^2
-k_3^2-{m^*_s}^2)}[W]_3 = \sigma(k_3)j_l(k_3r)Y_{l0}\; ,
\label{s3}
\end{equation}
\begin{equation}
[\delta V_0]_3={2g_v\overline{\epsilon}_F\overline{p}_F\over\pi^2
(\omega_n^2-k_3^2-{m}^2_v)}[W]_3=V_0(k_3)j_l(k_3r)Y_{l0}\; ,
\label{v3}
\end{equation}
\begin{equation}
[\delta V_i]_3={G_v(k_3)\over g_v}\left(f(k_3)-{2\overline{p}_F^2k_3^2
\over15}\right)\partial_i(j_l(k_3r)Y_{l0})\; ,
\end{equation}
\begin{equation}
\left.{x}_i[\delta R_i]_3=-{{x}_i\partial_i[\delta\sigma]_3
\over g_s\overline{\rho}_{s0}}\right|_{r=R_0}\; ,
\end{equation}
\end{mathletters}
plus the corresponding solutions to the canonically conjugated fields.
In the above equations we have introduced the functions
\begin{equation}
f(k)=-{2\overline{p}_F^4k^4(1+G_v(k))G_{0s}(k)\over15(3\overline
{\epsilon}_F^2\omega_n^2-\overline{p}_F^2k^2(1+G_v(k))G_{0s}(k))}
\; , \label{fk}
\end{equation}
and
\begin{equation}
G_{0s}(k)=1-{2\overline{p}_F\over\pi^2\overline{\epsilon}_F}
\left({g_v^2\overline{\epsilon}_F^2\over\omega_n^2-k^2-m_v^2}
-{g_s^2{M^*}^2\over\omega_n^2-k^2-{m^*_s}^2}\right)\; ,
\label{g0s}
\end{equation}
with $G_v(k)$ defined in Eq.(\ref{gv}).

The dispersion relation obeyed by these  solutions is
\begin{eqnarray}
& &3\overline{\epsilon}_F^2\omega_n^2\left({5\overline{\epsilon}_F^2
\omega_n^2\over\overline{p}_F^2}-{9k_3^2\over7}\right)(\omega_n^2
-k_3^2-{m_v^*}^2)(\omega_n^2-k_3^2-{m_s^*}^2)=
\nonumber\\*[7.2pt]
&=&9(\overline{\epsilon}_F^2\omega_n^2k_3^2-{\overline{p}_F^2k_3^4\over 7})
(\omega_n^2-k_3^2-m_v^2)(\omega_n^2-k_3^2-{m_s^*}^2)-
{2\overline{p}_F\over\pi^2\overline{\epsilon}_F}\left(5\overline
{\epsilon}_F^2\omega_n^2k_3^2\right.
\nonumber\\*[7.2pt]
&-&\left.{9\over7}\overline{p}_F^2k_3^4\right)
[g_v^2\overline{\epsilon}_F^2(\omega_n^2-k_3^2-{m_s^*}^2)
-g_s^2{M^*}^2(\omega_n^2-k_3^2-m_v^2)]\; .
\label{d3}
\end{eqnarray}
There are four solutions of kind 3, two more than the number of
this kind of solutions found in \cite{vos2}. This should  be expected
since, besides the vector meson field, the scalar meson field also
couples to the longitudinal solution $[\phi_{ij}]_3$. It is easy to show 
that for $g_s=0$ and $g_v=0$ one recovers the two solutions of
ref.\cite{vos2}. 

Therefore, the Walecka model leads to the appearance of 7 different
values for $k$ for a fixed frequency $\omega$, in contrast with
the model of ref.\cite{vos2}, which gives only 4 different values.

There is still a fourth kind of solution for the tensor fields, which
can be chosen to be
\begin{equation}
[\phi_{ij}]_4=[\chi_{ij}]_4=\delta_{ij}F(r)Y_{l0}\; ,
\end{equation}
coupled to the scalar fields
\begin{equation}
[W]_4=[\psi]_4=-{\overline{p}_F^2\over2}F(r)Y_{l0} \; ,
\end{equation}
and to the meson fields
\begin{equation}
[\delta\sigma]_4=[\delta V_0]_4=[\delta V_i]_4=0\; ,
\end{equation}
where $F(r)$ is an arbitrary function. This solution is not trivial
because of the boundary condition Eq.(\ref{bo7}).

The general solution, for each normal mode, is a linear combination
of the eight particular solutions:
\begin{equation}
\phi_{ij}^{(m)} = c_1[\phi_{ij}(k_1r)]_1+\sum_{n=1}^2 c_{2n}[
\phi_{ij}(k_{2n}r)]_2+\sum_{n=1}^4 c_{3n}[\phi_{ij}(k_{3n}r)]_3
+c_4[\phi_{ij}(r)]_4 \; ,
\label{geso}
\end{equation}
with similar expressions for the other fields.

To avoid  zero frequency modes linked to the surface motion,  
we introduce in the
model a surface energy which, in a classical approximation, is given by
\begin{equation}
E_{sup}^{(2)}={\sigma_{sup}\over2R_0^2}(l(l+1)-2)\int d{\mathbf\Sigma}.
\delta{\mathbf R}\,\delta{\mathbf R}.\hat n\; , 
\end{equation}
where $\sigma_{sup}$ is the surface tension coefficient. 
This term does not alter
the equations of motion and, therefore, the dispersion relations. 
It only changes the boundary condition Eq.(\ref{bo6}) to 
\begin{equation}
\left.\dot{\psi}+{\overline p_F^2\over10}\dot{\phi}_{ii}+g_v\delta V_0
-g_s{\overline\rho_{s0}\over\overline\rho_0}\delta\sigma
-{\sigma_{sup}\over2R_0^2 \rho_0}(l(l+1)-2) \hat n\cdot\delta {\mathbf R}
\right|_{r=R_0}=0\; .\label{bo6m}
\end{equation}

Using the general solutions in the boundary conditions Eqs.(\ref{bo1}) to
(\ref{bo5}), Eq.(\ref{bo7}) and Eq.(\ref{bo6m}) we get the equations
(\ref{c1}) to (\ref{c8}) given in the appendix.
The eigenvalues are determined by solving consistently the dispersion
relation equations, Eqs.(\ref{d1}), (\ref{d2}) and (\ref{d3}), subjected
to the boundary conditions.

\subsection{Sum Rules}

Sum rules can be regarded as a  test to the validity of a  particular 
 nuclear model. Suppose that
a nucleus is excited from its ground state $|0\rangle$ to an excited state
$|n\rangle$, with an energy $E_n$, due to interactions with
an external field. One can define momenta, weighted in energy, of
the excitation strength distribution
\begin{equation}
m_k=\sum_n(E_n-E_0)^k|\langle n|\hat O|0\rangle|^2 \; ,
\end{equation}
where $\hat O$ is the one-body hermitian operator, responsible for
the excitation. In the above expression, $k=0,\pm1,\pm2,...$ and $
|n\rangle$ stands for a set of eigenstates of the hamiltonian of the 
system. A sum rule is obtained when it is possible to relate a momentum
with a known quantity.

The energy weighted sum rule (EWSR), $m_1$, is obtained through the
calculation of the expectation value of a double commutator
\begin{equation}
m_1=\sum_n(E_n-E_0)|\langle n|\hat O|0\rangle|^2 ={1\over2}
\langle 0|[\hat O,[H,\hat O]]|0\rangle \; .
\end{equation}

In the present problem, the general solution for the variational fields
 is given by the real part of
\begin{equation}
\Psi({\mathbf x},t)=\sum_n a_n
\left(\begin{array}{c}
iW^{(n)}({\mathbf x})  \\ i\delta\sigma^{(n)}({\mathbf x})  
\\  i\delta V_0^{(n)}({\mathbf x}) \\ i\delta{\mathbf R}^{(n)}({\mathbf x}) 
\\ i{\mathbf\Pi}_v^{(n)}({\mathbf x}) \\ i\chi_{ij}^{(n)}({\mathbf x})
\\ \phi_{ij}^{(n)}({\mathbf x})
\\
\psi^{(n)}({\mathbf x}) \\ \Pi_\sigma^{(n)}({\mathbf x}) 
\\ \delta{\mathbf V}^{(n)}({\mathbf x})
\end{array}  \right) e^{-i\omega_nt} \;,
\end{equation}
where the coefficients $a_n$ are determined by the initial conditions.
In order to derive the EWSR for the electric modes we consider the
following initial condition
\begin{mathletters}
\begin{equation}
\psi({\mathbf x},0)=D({\mathbf x})\; ,
\label{ic1}
\end{equation}
\begin{eqnarray}
\phi_{ij}({\mathbf x},0)&=&\chi_{ij}({\mathbf x},0)=W({\mathbf x},0)=\delta
{\mathbf R}({\mathbf x},0)=\delta{\mathbf V}({\mathbf x},0)
\nonumber\\*[7.2pt]
&=&\delta\sigma({\mathbf x},0)=\delta V_0({\mathbf x},0)=\Pi_\sigma({\mathbf x},0)=
\Pi_{V_i}({\mathbf x},0)=0\; ,
\label{ic2}
\end{eqnarray}
\end{mathletters}
with $D({\mathbf x})$ to be specified. We then expand the fields $\psi
({\mathbf x},0)$, $\Pi_\sigma({\mathbf x},0)$, $\phi_{ij}({\mathbf x},0)$
and $\delta{\mathbf V}({\mathbf x},0)$ as $
\varphi({\mathbf x},0)=\sum_n a_n\varphi^{(n)}({\mathbf x})$,
where, from the orthogonality relation Eq.(\ref{norma}), we get
\begin{equation}
a_n=\int d^3x {2\epsilon_Fp_F\over\pi^2}\left(W^{(n)}+
{p_F^2\over6}\chi_{ii}^{(n)}\right)D({\mathbf x})+\overline{\rho}_0
\int d{\mathbf\Sigma}.\delta{\mathbf R}^{(n)}D({\mathbf x})\;. 
\end{equation}
The coefficients $a_n$ are related to the expectation value of the 
transition operator, $a_n=\sqrt{2}\langle{n|\hat O|0}\rangle$. 
Therefore, the EWSR can be written as
\begin{equation}
m_1=\sum_n|a_n|^2\omega_n=2 E^{(2)}
\; ,
\label{sr}
\end{equation}
and, for the initial condition given in Eq.(4.24), the EWSR reads
\begin{equation}
\sum_n |a_n|^2\omega_n = \int d^3x {\overline{\rho}_0\over\overline
{\epsilon}_F}{\mathbf\nabla} D.{\mathbf\nabla} D \; .
\end{equation}
\section{Numerical Results}
We have performed our calculations with two different sets of the 
mean-field values of the parameters in the Walecka model:
\begin{itemize}
\item[I.] $g^2_s = 122.88, g^2_v = 169.49, \overline{p}_F = 1.3 fm^{-1}$,
$M^*/M = 0.522$
\item[II.] $g^2_s = 91.64, g^2_v = 136.20, \overline{p}_F = 1.42 fm^{-1}$, 
$M^*/M = 0.556$
\end{itemize}
where $M=938$ MeV and $M^*$ is the effective mass.  The effective mass and 
the Fermi momentum indicated for each set correspond to the values at which
saturation  of nuclear matter is obtained with an 
energy per nucleon $E/N=-15.75$ MeV, using $m_s=550$ MeV and $m_v=783$ MeV.
The surface tension,
from the liquid drop model \cite{myers}, is 
$\sigma_{sup}=1.017$ MeV/fm$^2$.  The results were calculated for a 
nucleus with $A=208$. The radius $R_0$ is obtained from the value of
$\overline{p}_F$ corresponding to the chosen set of parameters.

For the  excitation operator introduced in   \ref{ic1}  we will  use 
\begin{mathletters}
\begin{eqnarray}
D({\mathbf x})&=&r^2Y_{00},\;\;\;\;\;\;\;\;\;\; l=0\; ,
\label{d0}\\*[7.2pt]
&=&r^l Y_{l0},\;\;\;\;\;\;\;\;\;\; l\geq2\; .
\label{dl}
\end{eqnarray}
\end{mathletters}

Table I shows  the energies of the normal modes together with the 
corresponding 
percentage of the exhausted energy weighted sum rule (EWSR), for 
the two sets given above and for different multipolarities. 
The EWSR is fragmented over the whole range of energies and
only the nuclear  modes which exhaust more than 0.1\% of the sum rule  
are given. The distribution of the EWSR between the nuclear 
modes and the mesonic modes (energies larger than the meson masses)
agrees with the results obtained in \cite{nos1}, 
where it is shown that in infinite nuclear matter and for small momentum 
transfer about 62\% of the EWSR 
is exhausted by the continuum nuclear modes and about 38\% by the vector 
meson modes. For instance, for $l=2^+$ and for set I, we find a vector
meson mode at $\hbar w_i=984.56$MeV which exhausts 27.30\% of the EWSR.
The other mesonic modes are not as collective as this one and are
distributed over a large range of energies. This pattern is reproduced
for the two sets of parameters and for all multipolarities. 
The EWSR is fulfilled considering all the nuclear and mesonic modes.
 In non-relativistic
calculations using the same nuclear fluid-dynamical model used here
\cite{vos1,vos2,joao}, the mesonic modes are not present and, therefore,
the EWSR is distributed only through the nuclear modes.
From this Table we can see that for set II the collective modes 
come at a slightly higher energy than in set I and that the strength is
more concentrated at higher energies. 

In table II we give for set I  and for $l^\pi=0^+, 
\, 2^+,\,  3^-,\, 4^+$ the energy of the normal modes 
with energy below 100 MeV (first column) and the 
corresponding percentage of the energy weighted sum rule (EWSR) (third 
column). In the second column we present a renormalized 
percentage of the EWSR,
 renormalizing the strength distributed among  states with energy below 130 
MeV to  1. The 
renormalizing factor is $m'_1(l)=0.56,\, 0.60,\, 0.61,\, 0. 56$,
  respectively, for $l=0, \, 2, \, 3, \,4$. 
This is 
done so that we can compare more easily the results obtained in the present 
work with previous results obtained in a non-relativistic fluid-dynamical 
model, \cite{joao},
(columns 4 and 5) and experimental data (columns 
6 and 7)\cite{exp}. 
 Looking at the modes with energy 
below 100 MeV,
we may immediately conclude that there is a certain correspondence between 
the   states obtained in the present approach  and the ones of  \cite{joao},
 if we identify the states by the percentage of the  
exhausted EWSR. However, the corresponding states come, in the present 
relativistic approach, at higher energies. For instance, the quadrupole
low lying mode and giant resonance  come, respectively,  at 10 and 20 MeV and 
exaust 8\% and 77\% of the EWSR 
while the experimental modes come at 4  and 11 MeV and exhaust 15\% and 70\% 
of the EWSR. Another possible way of identifying the modes is done by 
 comparing the current 
transition density (2.8) and the transition density (2.7) for these two 
modes with the ones of ref. \cite{joao}. 
In figure 1 and 2 we plot  $j_+,\, j_-, \, j_{div}$ (arbitrary units)
 defined by the equations: 
$${\mathbf j}({\mathbf r})= 
j_+({ r}) {\mathbf Y}_{l,l+1, 0}(\Omega)
+j_-({ r}) {\mathbf Y}_{l,l-1, 0}(\Omega),$$
$${\mathbf \nabla}\cdot {\mathbf j}({\mathbf r})= j_{div}(r) Y_{l 0}.$$
The function $j_{div}$ is related to the transition density $\delta \rho$ 
($\dot\rho=-{\mathbf \nabla}\cdot {\mathbf j}$). For the 10.03 MeV mode,
$j_+$  and $j_-$ have oposite signs and $j_{div}$ is close to zero, 
characteristic of a surface mode. These are 
typical properties of a low lying mode. For the 20.15 MeV mode, $j_+$ and 
$j_-$ have the same sign and $j_{div}$ comes diferent from zero for
$r/R_0 > 0.5$. This behaviour is closer to the behaviour expected from a 
giant resonance. 
We conclude the identification we have done is correct.

We note that our modes with the lowest energy have energies 
similar to the ones obtained  in ref. \cite{fuse}, 
however,  these are not 
the states that exhaust the largest percentage of the EWSR and, therefore,
they should not be identified with the giant resonances. 
The breathing mode comes at a very high energy, but this was expected
 owing to the high incompressibility of the model.
 
While in ref.\cite{fuse} only the lowest modes were determined, we
have found all the modes that exhaust a significant fraction of the
corresponding EWSR (which we also derived). Furthermore, we have shown
that the lowest modes are not the most collective ones. 

From the present results we conclude that the dynamical properties of the 
nuclei are not so well described by the Walecka model as the static 
properties 
such as densities and single particle energies. In our calculation 
we have taken for the ground-state of the nucleus a Slater determinant 
derived from a square-well instead of the
 the self-consistent ground-state.
We believe, however, that for 
large nuclei such as the $^{208}$Pb nucleus this is a good approximation which 
allows us to obtain analytical expressions for the equations of motion and 
the boundary conditions.  

\bigskip

\section{Appendix}
\bigskip

Using the general solutions in the boundary conditions Eqs.(\ref{bo1}) to
(\ref{bo5}), Eq.(\ref{bo7}) and Eq.(\ref{bo6m}) we get the following equations:
\begin{mathletters}
\begin{eqnarray}
& &(2-l(l+1))(r\partial_r+1)c_1j_l(k_1r)+\sum_{n=1}^2\left[2l(l+1)-2
r\partial_r-2-k_{2n}^2r^2\right]c_{2n}j_l(k_{2n}r)
\nonumber\\*[7.2pt]
& &+\left.\sum_{n=1}^4(r\partial_r-1)c_{3n}j_l(k_{3n}r)\right|_{r=R_0}=0
\label{c1}
\end{eqnarray}
\begin{eqnarray}
&&l(l+1)(2-l(l+1))c_1j_l(k_1r)+\sum_{n=1}^22l(l+1)(r\partial_r-1)
c_{2n}j_l(k_{2n}r)
\nonumber\\*[7.2pt]
& &+\left.\sum_{n=1}^4\left[l(l+1)-2r\partial_r-{2\over3}k_{2n}^2r^2
\right]c_{3n}j_l(k_{3n}r)+r^2c_4F(r)\right|_{r=R_0}=0 \; ,
\label{c2}
\end{eqnarray}
\begin{eqnarray}
& &2\left[k_1^2r^2(r\partial_r-1)+6(r\partial_r+1)-{l(l+1)\over2}r
\partial_r\right]c_1j_l(k_1r)-2\sum_{n=1}^2\left[6(r\partial_r+1)\right.
\nonumber\\*[7.2pt]
& &\left.\left.+2k_{2n}^2r^2-3l(l+1)\right]c_{2n}j_l(k_{2n}r)+3
\sum_{n=1}^4(r\partial_r-2)c_{3n}j_l(k_{3n}r)\right|_{r=R_0}=0 \; ,
\label{c3}
\end{eqnarray}
\begin{eqnarray}
& &l(l+1)\left[k_1^2r^2(r\partial_r-1)+3r\partial_r+12-3l(l+1)
\right]c_1j_l(k_1r)
\nonumber\\*[7.2pt]
& &+\sum_{n=1}^2l(l+1)\left({2\over5}
k_{2n}^2r^2+6r\partial_r-12\right)c_{2n}j_l(k_{2n}r)
\nonumber\\*[7.2pt]
& &\left.+\sum_{n=1}^4
\left[{3\over5}k_{3n}^2r^3\partial_r+3\left(-3r\partial_r+l(l+1)
-k_{3n}^2r^2\right)\right]c_{3n}j_l(k_{3n}r)\right|_{r=R_0}=0 \; ,
\label{c4}
\end{eqnarray}
\begin{equation}
\left.\sum_{n=1}^2G_v(k_{2n})k_{2n}^4r^4c_{2n}j_l(k_{2n}r)
\right|_{r=R_0}=0 \; ,
\label{c5}
\end{equation}
\begin{eqnarray}
& &\sum_{i=1}^2{\omega_nG_v(k_{2i})\over5g_v}\overline{p}_F^2
k_{2i}^2l(l+1)c_{2i}j_l(k_{2i}r)+\sum_{i=1}^4\left[V_0(k_{3i})\right.
\nonumber\\*[7.2pt]
& &\left.\left.-{\omega_nG_v(k_{3i})\over g_v}\left(f(k_{3i})-
{2\overline{p}_F^2k_{3i}^2\over15}\right)-{g_v\overline{\rho}_0\over
g_s\overline{\rho}_{s0}}\sigma(k_{3i})\right]r\partial_rc_{3i}
j_l(k_{3i}r)\right|_{r=R_0}=0 \; ,
\label{c6}
\end{eqnarray}
\begin{eqnarray}
& &\sum_{i=1}^2{\overline{p}_F^2k_{2i}^2\over5}l(l+1)(1+G_v(k_{2i}))
c_{2i}j_l(k_{2i}r)-\sum_{i=1}^4\left[(1+G_v(k_{3i}))\left(f(k_{3i})
\right.\right.
\nonumber\\*[7.2pt]
& &\left.\left.\left.-{2\overline{p}_F^2k_{3i}^2\over15}\right)-
{\overline{\epsilon}_F\omega_n\over g_s\overline{\rho}_{s0}}\right]
r\partial_rc_{3i}j_l(k_{3i}r)\right|_{r=R_0}=0 \; ,
\label{c7}
\end{eqnarray}
\begin{eqnarray}
& &\left.c_4F(r)\right|_{r=R_0}={5\over\omega_n\overline{p}_F^2}
\sum_{i=1}^4\left[\omega_nf(k_{3i})-g_vV_0(k_{3i})+{g_s
\overline{\rho}_{s0}\over\overline{\rho}_0}\sigma(k_{3i})\right.
\nonumber\\*[7.2pt]
& &\left.\left.-{\sigma_{sup}\over g_s\overline{\rho}_0\overline
{\rho}_{s0}R_0^2}(l(l+1)-2)\sigma(k_{3i})\partial_r\right]
c_{3i}j_l(k_{3i}r)\right|_{r=R_0}=0 \; ,
\label{c8}
\end{eqnarray}
\end{mathletters}
with the functions $\sigma(k)$ and $V_0(k)$ defined in Eqs.(\ref{s3})
and (\ref{v3}).

\section{Acknowledgments}
This work was partially supported by CNPq, JNICT and 
 Funda\c c\~ao Calouste Gulbenkian. MN acknowledges the warm hospitality
and congenial atmosphere provided by the Centro de F\'\i sica
Te\'orica of the University of Coimbra during her stay in Portugal.

%

%
\begin{table}
\caption{Energies and fractions of the energy-weighted sum rule
for different multipolarities and different sets of parameters.}
\end{table}  

\begin{table}
\caption{Comparison between the energies and fractions of the
energy-weighted sum rule obtained in the present work (first column), 
in [20] 
(second column) and experimental data (third column) [24].}
\end{table}  

\begin{figure}
\caption{ $j_+$ (full-line), $j_-$ (dashed-line) and $j_{div}$ 
(dash-dotted-line) in arbitrary units for the $l^\pi=2^+$ E=10.03 MeV mode.}
\end{figure}

\begin{figure}
\caption{ $j_+$ (full-line), $j_-$ (dashed-line) and $j_{div}$ 
(dash-dotted-line) in arbitrary units for the $l^\pi=2^+$ E=20.15 MeV mode.}
\end{figure}

\newpage
\begin{center}
Table I
\end{center}
\renewcommand{\arraystretch}{.8} 
\begin{center}
  \begin{tabular}{|c|c c|c c|}
  \hline
  $l^{\pi}_i$ & I && II &\\ 
  \hline
  & $\hbar \omega_i$(MeV) & $m_1(\%)$ & $\hbar \omega_i$(MeV) & $m_1(\%)$\\   
\hline
$0^+_1$ & 28.56 & 14.06 & 37.27 & 8.55    \\ 
$0^+_2$ & 35.50 & 27.46 & 46.12 &30.34   \\
$0^+_3$ & 50.95 &  2.30 & 61.59 & 2.13     \\
$0^+_4$ & 68.25 &  1.32 & 81.14 & 0.25     \\
$0^+_5$ & 71.29 &  5.94 & 86.73 & 8.59 \\
$0^+_6$ & 88.51 &  0.58 & 98.92 & 0.71   \\
$0^+_7$ &105.11 &  1.07 &133.51 & 3.22    \\
$0^+_8$ &107.87 &  2.57 & & \\
  \hline
  $2^+_1$ & 10.03 &  4.67  & 11.90 &  1.46\\ 
  $2^+_2$ & 20.15 & 45.32  & 28.07 & 42.51\\
  $2^+_3$ & 28.32 &  0.78  & 33.77 &  5.12\\
  $2^+_4$ & 35.32 &  3.95  & 39.94 &  0.48\\
  $2^+_5$ & 35.82 &  0.31  & 42.33 &  4.64\\
  $2^+_6$ & 49.35 &  0.52  &59.72 &  0.35\\
  $2^+_7$ & 64.20 &  1.93  & 75.59 &  0.96\\
  $2^+_8$ & 69.91 &  0.10  &       &      \\   
\hline
  $3^-_1$ & 12.93 & 11.42  & 15.86 & 0.32\\
  $3^-_2$ & 14.44 &  1.09  & 17.77 & 5.71\\
  $3^-_3$ & 32.84 & 35.48  & 42.14 &35.24\\
  $3^-_4$ & 37.20 &  0.70  & 44.95 & 4.09\\
  $3^-_5$ & 42.33 &  0.08  & 47.65 & 2.17\\
  $3^-_6$ & 45.03 &  3.90  & 51.98 & 7.00\\
  $3^-_7$ & 57.55 &  0.64  & 68.89 & 0.53\\
  \hline
  $4^+_1$ & 18.06 & 13.41  & 22.35 & 0.69\\
  $4^+_2$ & 20.46 &  1.37  & 24.93 & 7.21\\
  $4^+_3$ & 43.94 & 16.44  & 54.32 &10.32\\
  $4^+_4$ & 45.91 & 14.54  & 57.52 &20.27\\
  $4^+_5$ & 48.84 &  2.05  & 62.47 &11.42\\
  $4^+_6$ & 54.83 &  3.47  & 73.41 & 0.11\\
  $4^+_7$ & 65.46 &  0.68  & 77.68 & 0.64\\
  \hline
  \end{tabular}
\end{center}
\vfill
\eject
\begin{center}
Table II
\end{center}
\renewcommand{\arraystretch}{.7} 
\begin{center}
  \begin{tabular}{|c|c c c|c c|c c|}
  \hline
  $l^{\pi}_i$ &  & present &  & \cite{joao} & & experimental \cite{exp} & \\ 
  \hline
  & $\hbar \omega_i$(MeV) & $m_1(\%)/m'_1(l)$ & $m_1(\%)$ 
&  $\hbar \omega_i$(MeV) & $m_1(\%)$ &  $\hbar \omega_i$(MeV) & $m_1(\%)$ \\
\hline\rule[-2mm]{0mm}{7mm}
$0^+_1$ &     28.56  & 25.23 & 14.06    &  15.87 & 95.15 &  13.9 & 100. \\ 
$0^+_2$ &     35.50  & 49.28 & 27.46    &  18.95 &  2.26 & &\\
$0^+_3$ &     50.95  &  4.14 &  2.30    &  28.14 &  0.03 & &\\
$0^+_4$ &     68.25  &  2.37 &  1.32    &  36.83 &  0.03 & &\\
$0^+_5$ &     71.29  & 10.66 &  5.94    &  41.29 &  1.46 & &\\
$0^+_6$ &     88.51  &  1.03 &  0.58    &   & & &\\
total &           &  99.99   & 55.72    &  & 98.98& &\\
  \hline
  $2^+_1$ &  10.03 &  7.67 &  4.56 &  3.73 & 30.90& 4.09 & 15.\\ 
  $2^+_2$ &  20.15 & 76.99 & 45.79 & 11.70 & 64.19 & 10.9 $\pm$ 0.3 & 70.0\\
  $2^+_3$ &  28.32 &  1.29 &  0.77 & 17.45 &  2.17& &\\
  $2^+_4$ &  35.32 &  6.35 &  3.78 & 20.54 &  1.10& &\\
  $2^+_5$ &  35.82 &  0.53 &  0.31 & 21.12 &  1.00& &\\
  $2^+_6$ &  49.35 &  0.88 &  0.52 & 27.30 &  0.06& &\\
  $2^+_7$ &  64.20 &  3.15 &  1.87 &  &  & &\\
  $2^+_8$ &  69.91 &  0.17 &  0.10 &  &   & &\\ 
  $2^+_8$ &  87.43 &  0.29 &  0.17 &  &   & &\\ 
  total   &        & 99.88 & 59.46 &  & 99.32& & \\
  \hline
  $3^-_1$ & 12.93  & 18.74 & 11.43 &  2.92 & 34.10& 2.61 & 33.\\
  $3^-_2$ & 14.44  &  1.71 &  1.04 &  8.43 &  0.29&  &  \\
  $3^-_3$ & 32.84  & 58.70 & 35.79 & 18.53 & 43.44& 18.4 $\pm$ 0.8 &
36.\\
  $3^-_4$ & 37.20  &  1.15 &  0.70 & 22.80 & 10.88& 21.8 $\pm$ 0.8  & 27.\\
  $3^-_5$ & 45.03  &  6.06 &  3.69 & 26.87 &  5.18& &\\
  $3^-_6$ & 57.55  &  1.05 &  0.64 &  & & &\\
  $3^-_7$ & 78.43  &  2.96 &  1.81 &  & & &\\
  $3^-_8$ & 82.04  &  3.10 &  1.89 &  & & &\\
  $3^-_9$ & 95.87  &  0.46 &  0.28 &  & & &\\
  total &        & 100.00 & 61.00  &  & 97.64& &\\
  \hline
  $4^+_1$ &  18.06 & 23.10 & 13.06 &  4.51 & 34.10& 4.32 &\\
  $4^+_2$ &  20.46 &  2.35 &  1.33 & 12.26 &  2.05& 12.0$\pm$ 0.3 & 10 
$\pm$ 3\\
  $4^+_3$ &  43.94 & 29.21 & 16.51 & 23.36 & 22.39& &\\
  $4^+_4$ &  45.91 & 25.77 & 14.57 & 27.64 &  8.86& &\\
  $4^+_5$ &  48.84 &  3.62 &  2.05 & 29.67 & 17.10& &\\
  $4^+_6$ &  54.83 &  5.81 &  3.28 & 33.45 &  8.45& &\\
  $4^+_7$ &  65.46 &  1.20 &  0.68 & 35.38 &  4.18& &\\
  total  &       & 100.00 & 56.52  &  &97.13 & &\\
  \hline
  \end{tabular}
\end{center}


\begin{references}
\bibitem{wal1} B. D. Serot and J. D. Walecka, Adv. Nucl. Phys. {\bf 16},  
1 (1986).
\bibitem{serot} B. D. Serot, Rep. Prog. Phys. {\bf 55}, 1855 (1992).
\bibitem{ha} S. Hama, B. C. Clark, E. D. Cooper, H. S. Sherif and
R. L. Marcer, Phys. Rev. {\bf C41}, 2737 (1990).
\bibitem{wa} S. J. Walace, Ann. Rev. Nucl. Part. Sci., 
{\bf 37}, 267 (1987).
\bibitem{md} T. D. Cohen, R. J. Furnstahl and D. K. Griegel, Phys. Rev. 
Lett.{\bf 67}, 961 (1991); X. Jin, M. Nielsen, T. D. Cohen, R. J. Furnstahl 
and D. K. Griegel, Phys. Rev. {\bf C49}, 464 (1994).  
\bibitem{furn1} R. J. Furnstahl, C. E. Price and G. E. Walker, Phys. Rev. 
{\bf C36}, 2590 (1987).
\bibitem{wal2} C. J. Horowitz and J. D. Walecka, Nucl. Phys. {\bf A364}, 
429 (1981).
\bibitem{wal3} J. D. Walecka, Phys. Lett. {\bf 94B}, 293 (1980).
\bibitem{fuse} R. J. Furnstahl and B. D. Serot, Acta Phys. Polon. 
{\bf B16}, 875 (1985).
\bibitem{soni} V. Soni, Phys. Lett. {\bf 183B}, 91 (1987).
\bibitem{tom} T. D. Cohen, M. Banerjee and C.-Y. Ren, Phys. Rev. 
{\bf C36}, 1653 (1987).
\bibitem{perry} R. J. Perry, Phys. Lett. {\bf 199B}, 489 (1987).
\bibitem{furn2} R. J. Furnstahl and C. J. Horowitz, Nucl. Phys. {\bf A485}, 
632 (1988).
\bibitem{nos1} M. Nielsen, C. da Provid\^encia and J. da Provid\^encia, 
Phys. Rev. {\bf C44}, 209 (1991).
\bibitem{nos2} M. Nielsen, C. da Provid\^encia and J. da Provid\^encia, 
Phys. Rev. {\bf C47}, 200 (1993).
\bibitem{mat} T. Matsui, Nucl. Phys. {\bf A370}, 365 (1981).
\bibitem{ho} K. Lim and C. J. Horowitz, Nucl. Phys. {\bf A501}, 
729 (1989).
\bibitem{vos1} J. da Provid\^encia, L. Brito and C. da Provid\^encia , 
Nuovo Cimento {\bf 87}, 248 (1985).
\bibitem{vos2} L. Brito and C. da Provid\^encia , 
Phys. Rev. {\bf C32}, 2049 (1985).
\bibitem{joao} J. da Provid\^encia Jr., Nucl. Phys. {\bf A582},
23 (1995).
\bibitem{bla} J. P. Blaizot, Phys. Rep. {\bf 64C}, 171 (1980).
\bibitem{st} S. Stringari, Nucl. Phys. {\bf A325}, 199 (1979).
\bibitem{myers} W. D. Myers and W. J. Swiatecki, Ann. Phys.   {\bf 84}
  (1974) 186
\bibitem{exp}  
M. N. Harakeh, B. van Heyst, K. van der Borg and A. van der Woude, Nucl. 
Phys. {\bf A327} (1979) 373;
H. P. Morsch, M. Regge, P. Turek and C. Mayer-Boriche, Phys. Rev. Lett. {
\bf 45 } (1980) 337;
C. Djalali, N. Marty, M. Morlet and A. Willis, Nucl. Phys. {\bf 380} (1982) 
42;
B. Bonin et al., Nucl. Phys. {\bf A430} (1984) 349;
F. E. Bertrand et al., Phys. Rev {\bf C 34} (1986) 45;
T. Suomij\"arvi et al, Nucl. Phys. {\bf A491} (1989) 314;
R. Liguori Neto et al., Nucl. Phys. {\bf A560} (1993) 733.
\end{references}
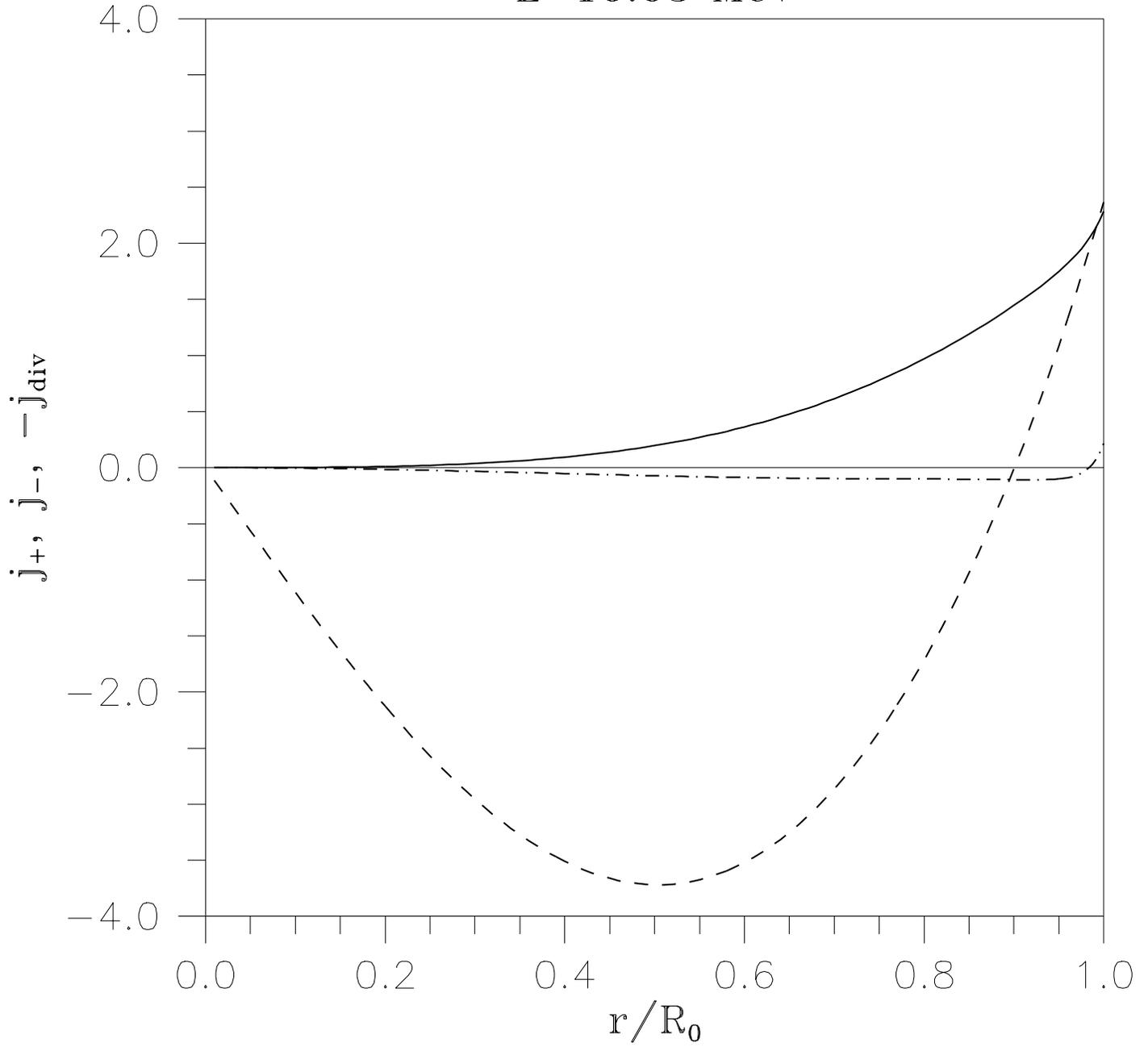
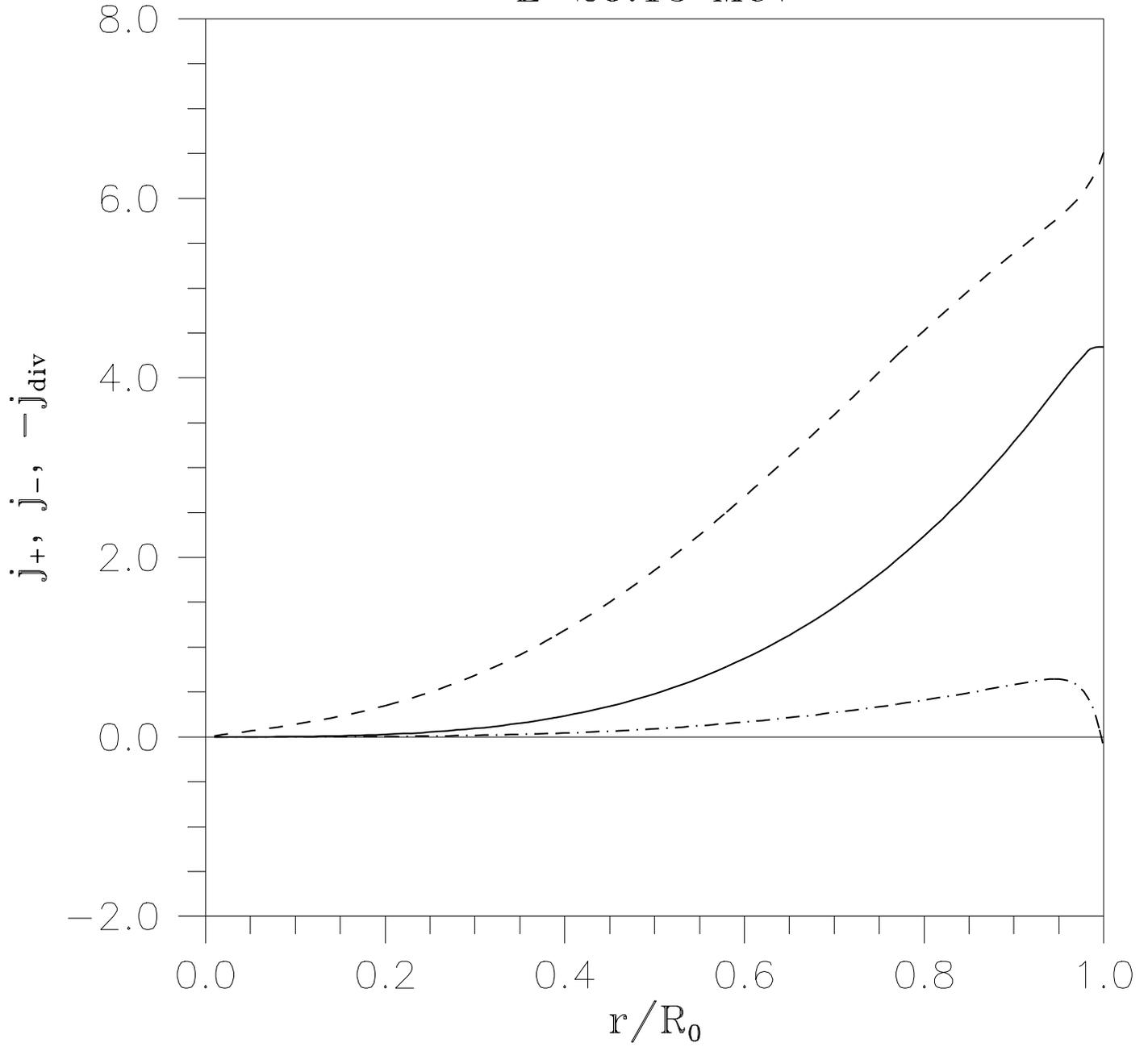
\end{document}